\documentclass[conference]{IEEEtran}
\IEEEoverridecommandlockouts
\usepackage{cite}
\usepackage{amsmath,amssymb,amsfonts}
\usepackage{algorithmic}
\usepackage{graphicx}
\usepackage{textcomp}
\usepackage{booktabs}
\usepackage{xcolor}
\def\BibTeX{{\rm B\kern-.05em{\sc i\kern-.025em b}\kern-.08em
    T\kern-.1667em\lower.7ex\hbox{E}\kern-.125emX}}

\usepackage{graphicx}
\usepackage{tikz}
\usetikzlibrary{arrows.meta,positioning,fit,shapes.geometric,shapes.misc,shapes.symbols,backgrounds}
\tikzset{
    component/.style={rectangle, rounded corners, draw=black, thick, align=center, font=\scriptsize, minimum width=2.3cm, minimum height=0.8cm},
    store/.style={cylinder, shape border rotate=90, draw=black, thick, aspect=0.25, minimum height=0.9cm, minimum width=1.4cm, font=\scriptsize},
    flow/.style={-Latex, very thin},
    ext/.style={},
    infra/.style={},
    sec/.style={}
}
\graphicspath{{media/}}
\usepackage{enumitem}

\begin{document}

\title{Evolution of AI Agent Registry Solutions: Centralized, Enterprise, and Distributed Approaches
}

\author{\IEEEauthorblockN{Aditi Singh}
\IEEEauthorblockA{\textit{Cleveland State University}\\
a.singh22@csuohio.edu}
\and

\IEEEauthorblockN{Abul Ehtesham}
\IEEEauthorblockA{\textit{Kent State University}\\
aehtesha@kent.edu}
\and

\IEEEauthorblockN{Mahesh Lambe}
\IEEEauthorblockA{\textit{Independent Researcher}\\
maheshlambe@gmail.com}
\and

\IEEEauthorblockN{Jared Grogan}
\IEEEauthorblockA{\textit{Independent Researcher}\\
grogan.jared@gmail.com}
\and

\IEEEauthorblockN{Abhishek Singh}
\IEEEauthorblockA{\textit{Massachusetts Institute of Technology}\\
abhi24@media.mit.edu}
\and

\IEEEauthorblockN{ Saket Kumar}
\IEEEauthorblockA{\textit{Northeastern University}\\
kumar.sak@northeastern.edu}
\and

\IEEEauthorblockN{ Luca Muscariello}
\IEEEauthorblockA{\textit{CISCO}\\
lumuscar@cisco.com}
\and

\IEEEauthorblockN{ Vijoy Pandey}
\IEEEauthorblockA{\textit{CISCO}\\
vijoy@cisco.com}
\and

\IEEEauthorblockN{ Guillaume Sauvage De Saint Marc}
\IEEEauthorblockA{\textit{CISCO}\\
g2stmarc@cisco.com}
\and
\IEEEauthorblockN{ Pradyumna Chari}
\IEEEauthorblockA{\textit{Massachusetts Institute of Technology}\\
pchari@mit.edu}
\and
\IEEEauthorblockN{ Ramesh Raskar}
\IEEEauthorblockA{\textit{Massachusetts Institute of Technology}\\
raskar@mit.edu}
}

\maketitle

\begin{abstract}
Autonomous AI agents now operate across cloud, enterprise, and decentralized domains, creating demand for registry infrastructures that enable trustworthy discovery, capability negotiation, and identity assurance. We analyze five prominent approaches: (1) MCP Registry (centralized publication of \texttt{mcp.json} descriptors), (2) A2A Agent Cards (decentralized self-describing JSON capability manifests), (3) AGNTCY Agent Directory Service (IPFS Kademlia DHT content routing extended for semantic taxonomy–based content discovery, OCI artifact storage, and Sigstore-backed integrity), (4) Microsoft Entra Agent ID (enterprise SaaS directory with policy and zero-trust integration), and (5) NANDA Index AgentFacts (cryptographically verifiable, privacy-preserving fact model with credentialed assertions). Using four evaluation dimensions—security, authentication, scalability, and maintainability—we surface architectural trade-offs between centralized control, enterprise governance, and distributed resilience. We conclude with design recommendations for an emerging Internet of AI Agents requiring verifiable identity, adaptive discovery flows, and interoperable capability semantics.
\end{abstract}


\section{Introduction}

Autonomous AI agents are rapidly becoming foundational across domains from cloud-native assistants and robotics to decentralized systems and edge-based IoT controllers. These agents act independently, make decisions, and collaborate at scale. As agent populations grow into the billions across heterogeneous platforms and administrative boundaries, the ability to \textit{identify, discover, and trust agents in real time} has emerged as a critical infrastructure challenge.

Traditional mechanisms like DNS and static service catalogs are poorly suited to agent ecosystems, which demand \textit{dynamic discovery}, \textit{verifiable metadata}, and \textit{privacy-preserving interactions}~\cite{li2023trustworthy}. Legacy systems assume fixed endpoints and ownership-based trust models, lacking the flexibility and cryptographic assurances needed for agents that rotate capabilities, change locations, and form ephemeral collaborations.

To address these limitations, several agent frameworks have introduced discovery metadata models. This paper focuses on three emerging approaches:

This paper presents a comparative analysis of five registry architectures:
\begin{itemize}[leftmargin=*]
    \item \textbf{MCP Registry}~\cite{modelcontextprotocolregistry}: centralized metadata publication for MCP servers via versioned \texttt{mcp.json}.
    \item \textbf{A2A Agent Cards}~\cite{a2aspecification}: decentralized JSON capability and endpoint description resolved at \texttt{/.well-known/agent.json}.
    \item \textbf{AGNTCY Agent Directory Service (ADS)}: IPFS Kademlia DHT content routing (extended for semantic taxonomy discovery), OCI artifact storage, Sigstore integrity proofs, and enterprise federation hooks.
    \item \textbf{Microsoft Entra Agent ID}~\cite{simons2025agentid}: managed enterprise directory integrating lifecycle, governance, conditional access, and policy enforcement.
    \item \textbf{NANDA Index: AgentFacts}~\cite{raskar2025upgrade}: decentralized verifiable credential (VC)-backed fact model supporting privacy-preserving dual-path discovery.
\end{itemize}

Rather than surveying general agent communication protocols~\cite{ehtesham2025surveyagentinteroperabilityprotocols}, this work is a focused comparison of these \textit{AI registry solutions}. It explores how each approach supports real-time discovery, identity validation, and cross-domain interoperability.

The paper is organized as follows:
Section II outlines background and motivation for agent registries. Section III introduces the evaluation framework and functional criteria. Sections IV–VI examine MCP, A2A, Microsoft Entra Agent ID and NANDA in depth. Section VII provides a comparative analysis across security, scalability, authentication, and maintainability. Section VIII concludes with design suggestions and practical recommendations for registry adoption in multi-agent systems.

By comparing these registries in the context of emerging agent infrastructure needs, this survey highlights the current gaps and emerging solutions driving the future of the Internet of AI Agents.

In trust-critical domains such as healthcare, finance, and critical 
infrastructure, agent registries must answer: \textit{How do we verify 
an agent is who it claims to be? How do we ensure its advertised 
capabilities haven't been tampered with? How do we revoke access when 
trust is violated?} This paper evaluates how five registry approaches 
address these trust primitives.

\section{Background}

The modern web operates on a reactive, client-driven model in which services wait for external requests before responding. Despite significant advances in cloud automation and event-driven design, this architecture remains largely inadequate for the emerging Internet of AI Agents , paradigm shift where autonomous, goal-directed software agents negotiate, coordinate, and act proactively on behalf of users.
Unlike traditional web resources, which are typically stateless and short-lived, autonomous AI agents are persistent computational entities capable of initiating control flow, retaining long-term memory, dynamically adapting to context, and spawning subordinate agents. These agents require infrastructure that supports high-frequency updates, real-time identity resolution, and trustable metadata exchange across heterogeneous systems and organizational boundaries.

This shift introduces significant challenges for discovery and coordination. The current Internet stack built on DNS, IP addressing, and certificate authorities was not designed to handle trillions of fast-moving, self-directed agents. Limitations in revocation latency, state propagation, identity verification, and routing scale all become critical bottlenecks, particularly 
in trust-sensitive domains requiring zero-trust architectures and continuous 
verification~\cite{ramezanpour2022intelligent, sedjelmaci2023zero}. The 
fundamental question of whether to upgrade existing infrastructure or implement 
purpose-built agent registries represents a qualitative, not incremental, 
architectural transition~\cite{raskar2025upgrade}.

To address these gaps, new registry models are emerging that shift away from static name-resolution systems to dynamic, metadata-rich discovery layers tailored to autonomous agents. This paper focuses on three such models, each coupled with a distinct metadata schema:

\begin{itemize}[leftmargin=*]
    \item \textbf{MCP Registry:} A centralized metaregistry that enables structured agent metadata publishing via \texttt{mcp.json}, supporting installability and versioning for MCP-compatible agents.
    \item \textbf{A2A Agent Cards:} A flexible, decentralized format for agent self-description, enabling discovery via well-known URLs, curated registries, or configuration files.
    \item \textbf{AGNTCY Agent Directory Service (ADS):} A content-addressed, OCI-aligned directory that resolves semantic capabilities to immutable digests and uses decentralized rendezvous (DHT) for provider discovery~\cite{agntcy_ads_spec}.
    \item \textbf{NANDA AgentFacts:} A cryptographically verifiable, privacy-preserving metadata schema designed for dynamic resolution, credentialed capability assertions, and federated environments.

\end{itemize}

These registry systems are positioned to address the foundational needs of the Internet of AI Agents: sub-second identity resolution, schema-validated capability representation, verifiable trust models, and privacy-aware discovery. Each model offers a different architectural stance, centralized, federated, or decentralized, on how to meet these requirements. This survey situates these three approaches within the broader transformation of web infrastructure, drawing historical parallels to transitions such as dial-up to broadband and IPv4 to IPv6. By understanding the limitations of existing systems and the unique demands of AI agents, we highlight why purpose-built registries are essential for scalable, secure, and interoperable agent ecosystems.

\subsection{Trust Requirements in Agent Registries}

Agent registries must provide three pillars of trust:

\begin{itemize}[leftmargin=*]
    \item \textbf{Identity Assurance}: Cryptographic binding between 
    agent identifiers and their metadata, preventing impersonation and 
    capability spoofing. Critical for healthcare agents handling 
    protected health information (PHI).

    \item \textbf{Integrity Verification}: Tamper-evident metadata and 
audit trails that detect unauthorized modifications~\cite{ashrafuzzaman2025blockchain}. 
Essential for regulatory compliance (HIPAA, FDA, GDPR).
    
    \item \textbf{Privacy Preservation}: Discovery mechanisms that don't 
    leak sensitive information about capabilities, access patterns, 
    or organizational relationships—particularly important in clinical 
    research and cross-institutional collaborations.
\end{itemize}

These informed our evaluation framework in Section~\ref{sec:comparison}.

\subsection{Design Evaluation Dimensions}

To compare candidate registry architectures against the above requirements, we evaluate along four core dimensions.

\begin{itemize}[leftmargin=*]
    \item  \textbf{Security:} Integrity of registry records and metadata via cryptographic signing. Resistance to spoofing, registry poisoning, and man-in-the-middle attacks.

    \item \textbf{Authentication:} Mechanisms for publisher identity verification (e.g., GitHub OAuth + DNS-TXT, DID-VC issuance, X.509 PKI). How registry updates are gated and how namespace ownership is enforced.

    \item \textbf{Scalability:} Ability to handle high lookup volumes and large agent populations via TTL-based caching, federated deployments, or CDN offload. Support for low-latency, geo-distributed resolution.

    \item \textbf{Maintenance:}  Operational simplicity: schema-first designs, minimal core code, decoupled metadata hosting.  Ease of upgrades, migration paths, and reduced patch surface by avoiding executable code hosting.

% \textbf{Governance:}
% Degree of community or multi-stakeholder control versus single-vendor operation.
% Transparency of policy, extensibility of namespaces, and processes for dispute and revocation.

   % \item  \textbf{Implementation Maturity (tie-breaker):} Readiness of the reference implementation or public service.  Availability of open-source code, live demos, and production deployments.
\end{itemize}

These dimensions provide a structured, source-grounded rubric for the comparative analysis in Sections 4–9.

\section{MCP Registry}\label{sec:mcp}

% \begin{figure}[!t]
%   \centering
%   \includegraphics[width=0.6\linewidth]{mcp_architecture.png}
%   \caption{Three‐layer architecture of the MCP Metaregistry: (1) Publisher Layer with CLI, GitHub OAuth and optional DNS TXT verification; (2) Registry Core Layer with Go REST API, Object Storage, MongoDB (and in‐memory fallback), and optional middle‐layers; (3) Consumer Layer with MCP client apps and mirrors.}
%   \label{fig:mcp_arch}
% \end{figure}

The MCP registry is a centralized “metaregistry” for discovering and installing MCP servers. Publishers push a versioned \texttt{mcp.json} via a CLI tool that performs a GitHub OAuth flow and, for reverse‐DNS namespaces, a DNS TXT challenge. The Go-based REST API exposes read endpoints (no authentication) and write endpoints (GitHub OAuth + DNS verification), stores raw JSON in object storage, indexes metadata in MongoDB (with an in-memory fallback), and generates asynchronous jobs or webhooks. Downstream MCP client apps poll the registry (or private mirrors), cache the data locally, and serve end-users without direct live calls to the central service.

\vspace{-3mm}
\subsection{Security}
The registry only accepts metadata from authenticated GitHub identities and, for domain-scoped namespaces, from DNS-verified domains. It does not host executable code; instead it holds metadata only, inheriting code-level security from established registries (npm, PyPI, DockerHub). This minimizes the attack surface and delegates authentication and domain control to proven systems.

\vspace{-3mm}
\subsection{Authentication}
All publish requests require a GitHub OAuth bearer token tied to the submitting user or organization. For reverse-DNS namespaces (e.g.\ \texttt{com.microsoft}), a DNS TXT record proof is required and linked to that GitHub identity. Read operations are openly accessible.

\vspace{-3mm}
\subsection{Scalability}
Only a small number of MCP client applications query the central registry; they cache and serve data to millions of end-users. The API supports asynchronous processing and webhooks. Metadata is stored in MongoDB suited to flexible, document-style records and served via CDN-cacheable HTTP endpoints; optional middle-layers (private mirrors, curated feeds) can shard load.

\vspace{-3mm}
\subsection{Maintenance}
The registry’s core service is schema-driven by \texttt{mcp.json} (OpenAPI/JSON Schema), with no package hosting or scanning to maintain. A CLI tool automates publication and verification flows. Schema updates proceed independently of the service code, and validation logic resides with the publisher.

% \begin{table}[!t]
%   \caption{MCP Metaregistry: Key Properties}
%   \label{tab:mcp_summary}
%   \centering
%   \begin{tabular}{@{}l p{11cm}@{}}
%     \toprule
%     \textbf{Dimension}    & \textbf{Summary} \\ \midrule
%     Security      & GitHub‐authenticated writes; DNS-TXT domain proof; metadata‐only storage. \\
%     Authentication & GitHub OAuth for publishing; DNS-TXT for reverse-DNS namespaces; open reads. \\
%     Scalability   & Client–only polling; CDN‐cacheable reads; MongoDB backend; optional webhooks. \\
%     Maintenance   & Schema‐driven \texttt{mcp.json}; no code hosting; CLI automation; decoupled schema evolution. \\
%     \bottomrule
%   \end{tabular}
% \end{table}

\section{Agent2Agent (A2A) Protocol}\label{sec:a2a}
The Agent2Agent (A2A) protocol is a transport-agnostic, enterprise-ready standard for inter-agent communication across heterogeneous systems. A2A enables autonomous agents potentially opaque, vendor-specific, or closed-source to discover, negotiate, and collaborate using a shared JSON-RPC interface over secure HTTP transport.

A2A is optimized for asynchronous, long-running, multimodal, and streaming interactions, supporting flexible task handoff between agents without requiring visibility into internal execution models. Through its declarative AgentCard, it enables dynamic discovery of skills, capabilities, and authentication requirements, establishing a standardized model for agent-to-agent collaboration.

% \begin{figure}
%     \centering
%     \includegraphics[width=\linewidth]{a2a.png}
%     % \includesvg[width=\linewidth]{a2a.svg}
%     \caption{An overview of A2A}
%     \label{fig:A2A}
% \end{figure}

\subsection{Security}
A2A relies on transport-layer security (TLS) and established web security best practices. Identity and authentication are handled outside of the A2A JSON-RPC payload via standard HTTP headers, allowing compatibility with OAuth2, API keys, and mTLS. Server identity is verified via TLS certificates, while clients authenticate based on security schemes advertised in the AgentCard. Push notifications (webhooks) are authenticated using per-client credentials, tokens, or schemes negotiated during setup. Agents do not share internal states; interactions are scoped to declared capabilities and managed through tasks and artifacts, reducing attack surfaces and limiting data exposure.

% \textbf{Summary:} Uses HTTPS + TLS for secure transport. Auth scheme discovery via AgentCard; supports OAuth2, mTLS, API keys. Tasks can enter auth-required state for dynamic in-task authentication. Push webhook security includes sender/receiver auth and tokens.

\subsection{Authentication}
AgentCards explicitly declare supported authentication mechanisms using OpenAPI-style security schemes (e.g., Bearer tokens, OpenID Connect, API keys). Clients must obtain credentials out-of-band and include them in request headers. Each RPC call is authenticated individually, and servers return HTTP 401/403 responses with guidance when credentials are missing or invalid. During execution, if secondary credentials are needed (e.g., to proxy tool access), tasks transition to auth-required, and clients supply the required credentials in subsequent messages.

% \textbf{Summary:} Authentication handled via standard HTTP headers. AgentCard includes discovery of required auth schemes. Supports dynamic credential escalation during task execution.

\subsection{Scalability}
A2A’s task-based, stateless transport over HTTP and SSE enables horizontal scalability across distributed agent systems. Agents define capabilities declaratively in AgentCards, allowing registries and discovery services to dynamically catalog available services. Tasks are long-lived objects with unique IDs, status updates, and artifact streams. Streaming (via SSE) and push notifications reduce polling overhead. Task lifecycles support fine-grained eventing (submitted, working, input-required, etc.), enabling responsive and resilient orchestration even in failure-prone environments.

% \textbf{Summary:} Stateless transport, long-lived task objects. Scalable eventing via SSE and webhooks. Agents can be discovered dynamically via .well-known/agent.json or registries.

\subsection{Maintainability}
A2A is intentionally simple and extensible, built atop HTTP and JSON-RPC 2.0. It minimizes custom logic and avoids bespoke protocols. Features like AgentCard and structured data formats (e.g., TextPart, DataPart, FilePart) ensure consistent interpretation of messages while allowing modality diversity. Schema evolution is flexible: agents can define capabilities per skill, override input/output MIME types, and extend their AgentCard dynamically. Task handling and message structure follow consistent, extensible conventions.

\section{AGNTCY Agent Directory Service}\label{sec:agntcy}
The AGNTCY Agent Directory Service (ADS) is organized around four
composable layers that separate data modeling, storage, distribution, and
content selection so the system can scale and evolve independently at each
concern:
\begin{enumerate}[leftmargin=*]
    \item \textbf{Agent AI metadata modeling (Open Agent Schema Framework, OASF).} Open Agent Schema Framework (OASF) cleanly separates what an agent can do from the conditions under which it may do it.
    \emph{Semantic attributes} describe the capability surface—skills and
    operational domains. \emph{Constraint descriptors} capture required
    dependencies (models, tools, data sources), resource and latency cost
    profiles, token budgets, and any compliance gates. This separation lets a
    resolver align intent with capability while enforcing execution bounds.
    \item \textbf{Immutable object storage (OCI).} Directory records
    are packaged as OCI-compliant artifacts and pushed to registries; each
    artifact is addressed by a cryptographic digest that enforces immutability
    and enables de\mbox{-}duplication. The OCI naming model yields globally
    unique references and efficient layer reuse for transport and caching.
    Human\mbox{-}readable aliases (tags or descriptive name attributes) can be
    attached without weakening the canonical digest binding.
    \item \textbf{Content distribution (OCI Distribution Spec).}
    Global replication, caching, and mirroring leverage the standard OCI
    Distribution Specification~\cite{oci_distribution_spec}, allowing any
    compliant registry to host identical artifacts (same digest) without
    coordination—enabling multi-source failover and locality-aware retrieval.
    \item \textbf{Content and registry selection (IPFS Kademlia DHT).}
    Selection reconciles heterogeneous registries via a two–step mapping: (1)
    semantic taxonomies (skills, domains) resolve to canonical OCI content
    identifiers (digests) through capability→digest indices; (2) each digest is
    mapped to one or more registry identifiers and locators (endpoints, tags,
    replica metadata). The extended Kademlia DHT~\cite{ipfs_kademlia} stores and
    routes both mapping layers, letting the content scheduler first derive the
    immutable content address, then choose optimal registries/replicas using
    locality and freshness cues. Once a digest is resolved and its minimal
    record (plus referenced artifact metadata) retrieved, rich filtering across
    the full OASF attribute set occurs locally; the DHT acts only as a
    rendezvous layer optimized for sparse semantic capability mappings rather
    than exhaustive attribute indexing.
\end{enumerate}

\paragraph{On Naming and Zooko's Triangle.}
By anchoring human-readable attributes to cryptographic OCI digests
(via signed OASF records) and resolving them through a decentralized DHT, ADS
satisfies the three traditionally conflicting properties of Zooko's triangle:
(1) decentralized naming, (2) secure naming (Sigstore provenance plus signed
attestations~\cite{agntcy_ads_draft}), and (3) human-readable naming (scoped
semantic attributes and constraints). A common criticism is that the ``secure''
leg, when implemented solely with Sigstore, still relies on a partially
centralized trust root. To mitigate this, ADS can augment Sigstore with a
trustless, smart-contract–anchored attestation path (e.g.,
ERC-8004~\cite{erc8004}) that commits record signature digests on-chain.

\subsection*{Architecture}
Building on the four layers above, ADS introduces a semantic
enrichment path. Hierarchical skill taxonomies accelerate capability-scoped
queries before DHT resolution. Replica metadata (latency, region, freshness)
feeds a composite scoring function that produces
deterministic, low-cost selection decisions. Location-independent OCI digests
(\texttt{sha256:<digest>}) enable transparent multi-registry reconciliation:
any registry serving the same digest becomes an interchangeable source. Digest
equality, reinforced by Sigstore provenance attestations,
enforces immutability and prevents substitution. Dynamic selection avoids
centralized choke points by coupling semantic relevance with Kademlia routing
proximity.

% ================= New Flow Diagrams (Publish + Discover) ==================
\begin{figure*}[!t]
    \centering
    % ================== Sequence (Time-Flow) Diagrams ==================
    % Shared style definitions local to figure
    \begin{tikzpicture}[%
        >=Latex,
        lifeline/.style={draw=black!60, thick},
        actor/.style={font=\scriptsize, align=center, text width=2.2cm},
        call/.style={-Latex, thin},
        note/.style={font=\tiny, align=left, text width=2.9cm},
        dashedcall/.style={-Latex, dashed, thin}
    ]
        % X positions for components (ordered as requested)
        \coordinate (pubX) at (0,0);
        \coordinate (apiX) at (2.8,0);
        \coordinate (localX) at (5.6,0);
        \coordinate (ociX) at (8.4,0);
        \coordinate (globalX) at (11.2,0);
        \coordinate (selectX) at (14.0,0);

        % Actor labels (shifted upward to avoid overlap with first messages)
        \node[actor] at ([yshift=0.85cm]pubX) {Publisher};
        \node[actor] at ([yshift=0.85cm]apiX) {Agent Directory\\API};
        \node[actor] at ([yshift=0.85cm]localX) {Local Search\\(Client)};
        \node[actor] at ([yshift=0.85cm]ociX) {OCI Storage};
        \node[actor] at ([yshift=0.85cm]globalX) {Global Search\\(DHT / Index)};
        \node[actor] at ([yshift=0.85cm]selectX) {Content \\ \& Remote\\Selection};

            % Lifelines (extend downward; total height ~12.5 to cover both flows)
            \foreach \x in {pubX,apiX,localX,ociX,globalX,selectX} {
                \draw[lifeline] (\x) -- ++(0,-12.5);
            }

            % ---------------- (a) Publishing Flow (upper half) ----------------
            % Shift entire publishing sequence downward to avoid label overlap
            % 1. Build & push artifact
            \draw[call] ([yshift=-0.6cm]pubX) -- node[above, sloped, font=\tiny]{Push artifact (digest)} ([yshift=-0.6cm]ociX);
            % 2. Register minimal record
            \draw[call] ([yshift=-1.6cm]pubX) -- node[above, sloped, font=\tiny]{Submit minimal OASF record} ([yshift=-1.6cm]apiX);
            % 3. API inserts semantic mapping to global search
            \draw[call] ([yshift=-2.6cm]apiX) -- node[above, sloped, font=\tiny]{Skill $\to$ digest mapping} ([yshift=-2.6cm]globalX);
            % 4. API registers digest $\to$ registry endpoints with selection service
            \draw[call] ([yshift=-3.6cm]apiX) -- node[above, sloped, font=\tiny]{Digest $\to$ replica endpoints} ([yshift=-3.6cm]selectX);
            % 5. Selection service acknowledges (optional)
            \draw[dashedcall] ([yshift=-4.6cm]selectX) -- node[above, sloped, font=\tiny]{ACK / status} ([yshift=-4.6cm]apiX);
            % 6. API acknowledgment to publisher
            \draw[dashedcall] ([yshift=-5.6cm]apiX) -- node[above, sloped, font=\tiny]{Publish complete} ([yshift=-5.6cm]pubX);

            % Bracket or label for publishing section
            \node[note] at (7.0,-6.1) {(a) Publishing Flow: artifact pushed first; minimal record + sparse semantic mapping + replica endpoints inserted.};

            % ---------------- (b) Discovery Flow (lower half) ----------------
            % Start discovery later to preserve white space
            % 1. Local capability or intent query -> global search
            \draw[call] ([yshift=-7.2cm]localX) -- node[above, sloped, font=\tiny]{Capability / embedding query} ([yshift=-7.2cm]globalX);
            % 2. Global search returns candidate digests (semantic rendezvous)
            \draw[dashedcall] ([yshift=-8.2cm]globalX) -- node[above, sloped, font=\tiny]{Candidate digests} ([yshift=-8.2cm]localX);
            % 3. Local fetch minimal records via API (may batch)
            \draw[call] ([yshift=-9.2cm]localX) -- node[above, sloped, font=\tiny]{Fetch minimal records} ([yshift=-9.2cm]apiX);
            % 4. API queries selection service for endpoints
            \draw[call] ([yshift=-10.2cm]apiX) -- node[above, sloped, font=\tiny]{Resolve digest $\to$ endpoints} ([yshift=-10.2cm]selectX);
            % 5. Selection returns ranked endpoints
            \draw[dashedcall] ([yshift=-11.2cm]selectX) -- node[above, sloped, font=\tiny]{Ranked endpoints} ([yshift=-11.2cm]apiX);
            % 6. API returns records + endpoints to local search
            \draw[dashedcall] ([yshift=-12.2cm]apiX) -- node[above, sloped, font=\tiny]{Records + endpoints} ([yshift=-12.2cm]localX);
            % 7. Local search pulls selected artifact
            \draw[call] ([yshift=-13.2cm]localX) -- node[above, sloped, font=\tiny]{Pull artifact (digest)} ([yshift=-13.2cm]ociX);

            % Discovery annotation
            \node[note, anchor=west] at (0.1,-14.1) {(b) Discovery Flow: DHT/global search supplies only sparse skill $\to$ digest; rich filtering happens locally using retrieved minimal records.};

    \end{tikzpicture}
    \caption{Time-flow (sequence) diagrams for AGNTCY Agent Directory. Vertical lifelines represent: Publisher, Agent Directory API, Local Search, OCI Storage, Global Search (semantic DHT / embedding index), and Content \& Remote Server Selection. (a) Publishing: artifact push precedes minimal record publication; sparse semantic (skill $\to$ digest) plus digest $\to$ endpoint mappings are inserted. (b) Discovery: local capability intent resolves via global search to digests, minimal records are fetched, local filtering/ranking occurs, endpoints resolved, and the selected artifact is pulled.}
    \label{fig:ads_timeflows}
\end{figure*}
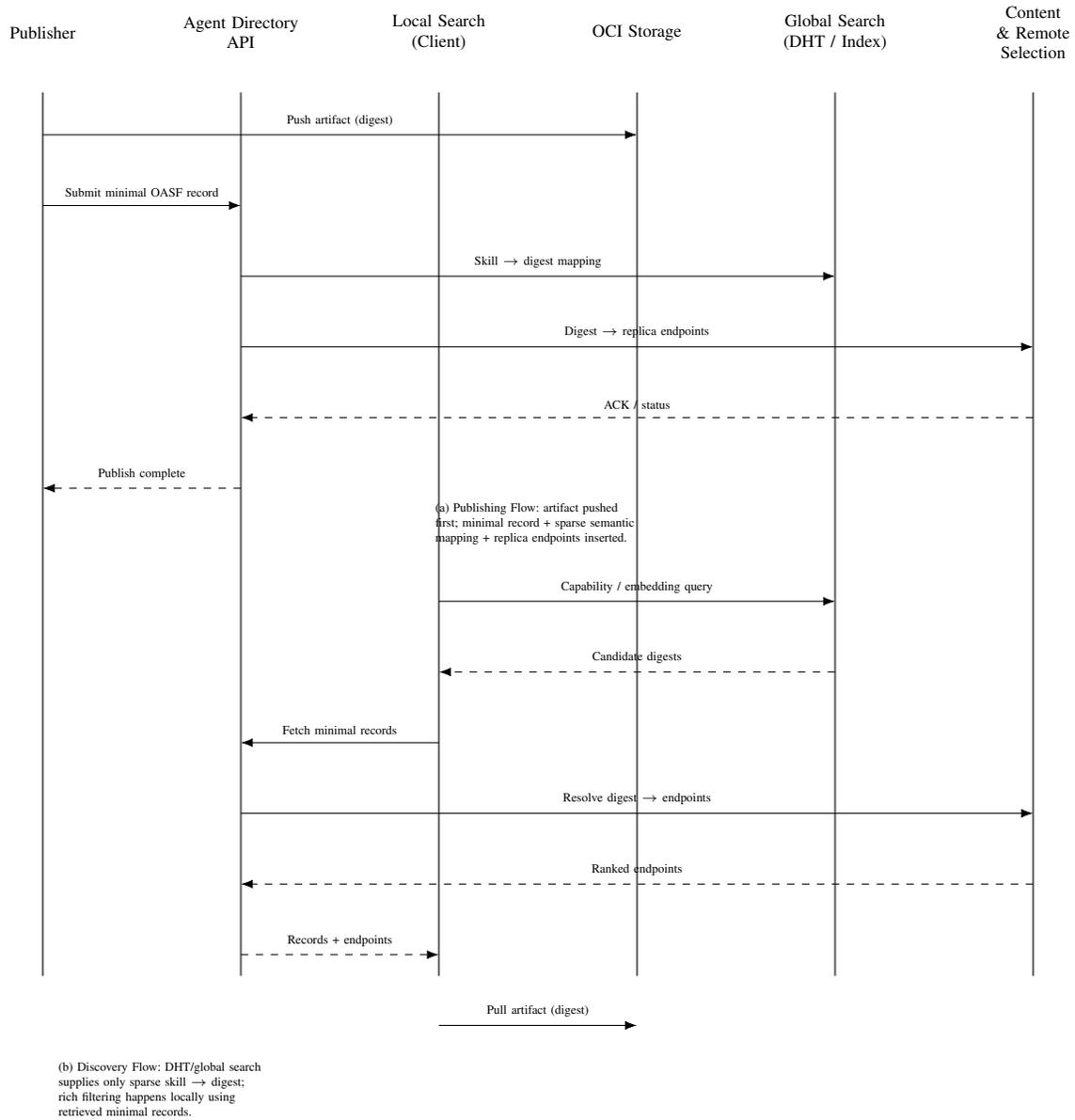

\subsection{Security} ADS security separates (1)
\textit{content / artifact security} from (2) \textit{server / runtime
security}, allowing strong integrity and provenance guarantees even in minimally
trusted network environments.

\subsection{Authentication}
Authentication in AGNTCY ADS is implemented in two distinct parts: data authenticity and server authentication.

\textbf{Data Authenticity.} This ensures that the data producer is authenticated and the integrity of the data is verifiable. All directory records are packaged as OCI records whose cryptographic digests are self-authenticating: the name (\texttt{sha256:<digest>}) is a direct function of the bytes, giving immutable, collision-resistant identifiers and secure naming by construction. Each published digest is accompanied by Sigstore provenance and signature materials (provenance attestation, record, producer identity, source revision), enabling pre-execution policy such as “only activate if provenance + record verified” without trusting a central admission service. Capability and constraint descriptors in the minimal OASF record are signed and bound (via the referenced digest) to the record they describe, preventing capability re-binding or privilege escalation through manifest substitution.

\textbf{Server Authentication.} This ensures trust among server peers and restricts unauthorized access. Core discovery can be left open (content trust is intrinsic), but operators often restrict which peers a directory node will accept or exchange routing/index data with. ``Secure mode'' introduces: (1) authenticated peer admission (static allowlist or dynamic attestation) before Kademlia session establishment; (2) mutual TLS with short-lived SPIFFE/SPIRE SVIDs or equivalent rotating credentials. These measures shrink the unsolicited attack surface (e.g., DDoS amplification/state exhaustion) and satisfy enterprise policies that forbid arbitrary federation. Such zero-trust peer authentication 
patterns are increasingly required in regulated 
environments~\cite{sedjelmaci2023zero}.

\subsection{Scalability} Every directory node contributes
routing, semantic index shards, and OCI artifact cache capacity. Adaptive
replication promotes hot capability shards and frequently requested artifact
digests near demand clusters. IPFS Kademlia iterative lookups (extended with
locality + semantic relevance scoring) are augmented by locality-aware peer
selection (latency + load) while opportunistic caching shortens future paths.
Joint optimization of OCI naming and distribution means that the same digest can
be resolved to the lowest-latency verified replica; if replicas diverge the
digest changes, forcing explicit revalidation. Content prefetch and layer
deduplication (OCI) reduces bandwidth while preserving deterministic digest
integrity.

\subsection{Maintainability} Declarative deployment
(Kubernetes with Helm) together with GitOps workflows minimizes control-plane
drift.  Cross-registry reconciliation through digest equality lets autonomous
registries evolve naming and tagging conventions independently while remaining
interoperable at the content-address level. Interoperability with external
registries is a distinctive feature: data from sources such as MCP or A2A can be
reconciled provided it is modeled into OASF via the corresponding module.
Modules for MCP and A2A are currently available.

\subsection{Enterprise Deployment Considerations}
The NANDA architecture has been explored in enterprise 
environments~\cite{wang2025using}, demonstrating cross-protocol 
interoperability with MCP, A2A, and standard HTTPS communications. The 
framework proposes Zero Trust Agentic Access (ZTAA) principles, extending 
traditional Zero Trust Network Access to address autonomous agent security 
challenges across heterogeneous protocol environments. This deployment 
experience informs the design choices outlined above, particularly regarding 
federation, policy enforcement, and credential management at scale.

\section{Microsoft Entra Agent ID}\label{sec:entra}

% \begin{figure}[!t]
%   \centering
%   %\includegraphics[width=\linewidth]{entra_agent_id_architecture.png}
%   \caption{Overview of Microsoft Entra Agent ID: unified directory of AI agent identities across Copilot Studio and Azure AI Foundry; enterprise enrollment, conditional access, and governance modules.}
%   \label{fig:entra_arch}
% \end{figure}

Microsoft Entra Agent ID provides a managed, enterprise-grade directory for AI agent identities. Agents created in Copilot Studio or Azure AI Foundry automatically appear as “Agent ID” applications in the Entra admin center. Identity practitioners gain visibility, lifecycle management, and access governance for these non-human identities using the same tools and policies as for user or service identities. Upcoming features include least-privilege token issuance, expanded Conditional Access, and cross-tenant identity federation. Further analysis of Microsoft Entra Agent ID’s security, authentication, scalability, and maintainability will be possible once technical documentation and operational data are available.

\begin{figure}[h]
    \centering
    \includegraphics[width=\linewidth]{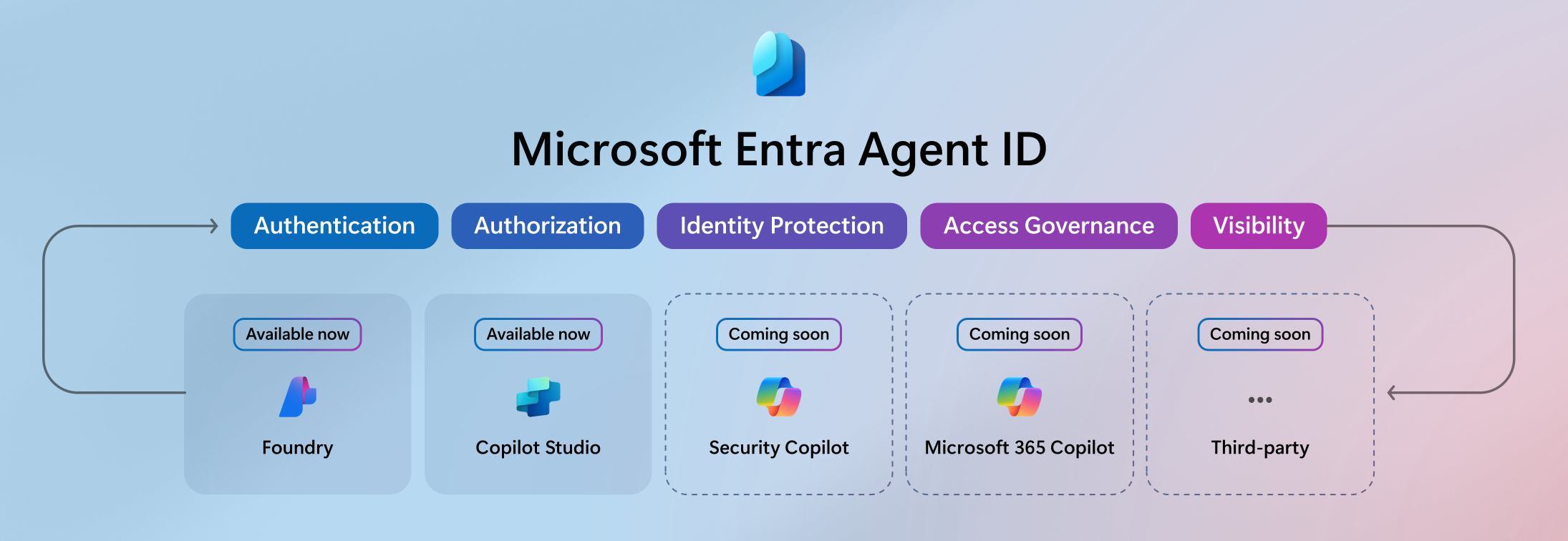}
    \caption{Microsoft Entra Agent ID Overview \cite{simons2025agentid}}
    \label{fig:microsoftentra_arch}
\end{figure}

\section{NANDA Index}\label{sec:nanda}

The Networked Agents and Decentralized AI (NANDA) Index, envisioned as a quilt of agents, resources, and tools registries spanning platforms, organizations, and protocols, presents a lean, modular architecture for agent discovery in decentralized environments. This quilt-like index supports the inclusion of both NANDA-native agents and third-party agents, enabling broad discoverability via a unified framework. Designed for scale, privacy, and interoperability, NANDA separates static identifier resolution from dynamic agent metadata to support rapid discovery, credentialed verification, and flexible routing across federated agent ecosystems.

Through this approach, NANDA index enables global interoperability, discoverability, and adaptable governance of agents without enforcing centralized control. Rather than acting as a universal authority, NANDA index allows commercial, governmental, and individual stakeholders to choose how their agents interact with the index. Agents can be listed directly within NANDA index or simply referenced via redirects to external platforms. This ensures that entities retain full control over access policies, trust management, and certification processes within their own domains, while still participating in a globally connected AI ecosystem.

At the core of the design is the concept of a minimal AgentAddr record an $Ed25519$ signed object that maps agent identifiers to one or more verifiable metadata locations: a public FactsURL, an optional privacy-preserving PrivateFactsURL, and an AdaptiveRouterURL for real-time routing. These records are lightweight ($\leq120$ bytes), cacheable, and stable, minimizing registry writes even in high-churn environments.

The full discovery flow is structured across three modular layers, each optimized for a specific role in scalable, decentralized agent ecosystems:

\begin{enumerate}[leftmargin=*]
    \item \textbf{Lean Index Layer}: Provides a decentralized, cacheable mapping from agent identifiers to signed \texttt{AgentAddr} records ($\leq120$ bytes), which include metadata URLs, cryptographic signatures, and TTLs. These lightweight, tamper-resistant records serve as the immutable anchor for discovery and routing without requiring frequent writes to the index.

    \item \textbf{AgentFacts Layer}: Distributes rich, schema-validated metadata using self-describing JSON-LD documents signed as W3C Verifiable Credentials. These documents describe capabilities, endpoints, and authentication logic, and can be hosted at agent-controlled domains or decentralized storage (e.g., IPFS), allowing privacy-preserving updates independent of the index.

    \item \textbf{Dynamic Resolution Layer}: Interprets metadata to route queries through static, rotating, or adaptive endpoints~\cite{zinky2025nanda}. This layer supports context-aware, real-time endpoint selection based on geographic location, system load, agent capabilities, and security threats. The AdaptiveResolver architecture enables short-lived endpoint rotation and negotiation of trust, quality of service, and resource constraints to meet performance, privacy, and resiliency requirements at scale.

\end{enumerate}

% The full discovery flow is distributed across three layers:
% \begin{enumerate}
%     \item \textbf{A Registry Layer} that serves as a static index of agent identifiers to metadata pointers.
%     \item \textbf{A Metadata Layer (AgentFacts)} containing schema-validated, W3C Verifiable Credential (VC)-signed descriptions of capabilities, endpoints, telemetry, and trust assertions.
%     \item \textbf{A Dynamic Resolution Layer} that supports static, rotating, or adaptive endpoint selection based on TTLs, load, geography, or policy.
% \end{enumerate}

\begin{figure}[!t]
  \centering
  \includegraphics[width=\linewidth]{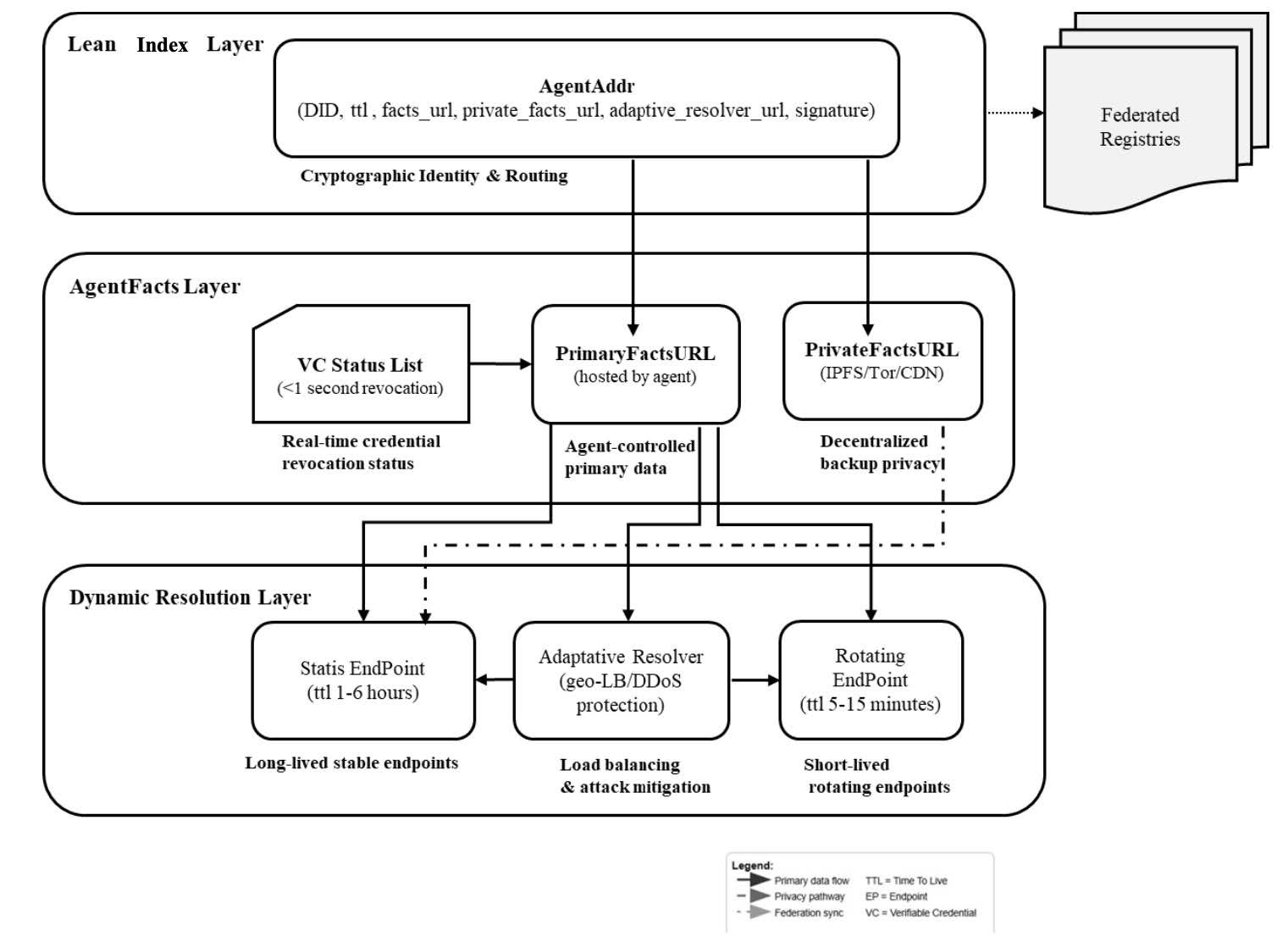}
    \caption{NANDA Index and AgentFacts Architecture: A modular three-layer system for decentralized AI agent discovery and routing.
  The \textbf{Lean Index Layer} resolves agent identifiers into signed \texttt{AgentAddr} records containing cryptographic identity, metadata URLs, and routing information, federated across registries.
  The \textbf{AgentFacts Layer} distributes dynamic, agent-controlled metadata via primary and private URLs, with real-time verifiable credential (VC) status updates for revocation.
  The \textbf{Dynamic Resolution Layer} enables endpoint resolution through stable, adaptive, and rotating strategies to support privacy, load balancing, and DDoS resilience.
  This layered design enables scalable, privacy-respecting, and interoperable agent discovery across federated domains~\cite{raskar2025dnsunlockinginternetai}.}
\label{fig:nanda_arch}
\end{figure}
\subsection{Security}
NANDA ensures security through end-to-end cryptographic guarantees. Each \texttt{AgentAddr} is signed using $Ed25519$ by the originating registry, ensuring authenticity, integrity, and immutability. AgentFacts documents are signed as W3C Verifiable Credentials (VCs v2), supporting short-lived credentials (often \textless5 minutes) and real-time revocation via VC-Status Lists. This cryptographic framework prevents identity spoofing, capability forgery, or impersonation across agent interactions. Additionally, the use of \texttt{PrivateFactsURL} supports privacy-preserving resolution pathways, helping obfuscate client access patterns in alignment with zero-trust security principles.

% NANDA achieves security through end-to-end cryptographic guarantees. Each AgentAddr is signed by the issuing registry to ensure authenticity and immutability. AgentFacts documents are signed by trusted issuers using W3C Verifiable Credentials v2 and support short-lived credentials (\textless 5 minutes) with revocation via VC-Status lists. This cryptographic framework ensures agents cannot spoof identities, impersonate others, or falsify capabilities. The inclusion of privacy-preserving resolution paths (PrivateFactsURL) further shields client access patterns from agents or intermediaries, in line with zero-trust design principles.

\subsection{Authentication}

Agent metadata is authenticated via decentralized identifiers (DIDs), and all claims within AgentFacts such as capabilities, compliance assertions, or audit results must be signed by credential authorities anchored to verifiable trust domains. Metadata updates do not require modifications to the Lean Index Layer; instead, they are independently hosted and signed by agents or authorized third-party infrastructure. This supports decentralized publication with verifiable provenance.

The Lean Index Layer does not mediate live authentication. Instead, it serves as a cryptographic root of trust, providing immutable references for downstream verification. Agent publishers may authenticate using DID-based credentials, issuer attestations, or delegated authority models supporting both self-sovereign and enterprise-aligned agent lifecycles.

\subsection{Scalability}
NANDA is architected for internet-scale performance through its layered and modular design. The Lean Index Layer minimizes write overhead by storing only signed pointers (\texttt{AgentAddr}), while metadata and operational state are managed externally in the AgentFacts and Dynamic Resolution layers. This separation eliminates write amplification and enables horizontal scaling.

TTL-based caching ensures that resolution requests are handled efficiently by edge resolvers or clients, reducing pressure on the Lean Index Layer. Meanwhile, the use of \texttt{AdaptiveResolverURL} and rotating endpoint pools enables responsive, real-time routing without burdening the index infrastructure.

The system supports federated deployment, where Lean Index instances can be domain-specific (e.g., healthcare, finance) or geo-localized. Cross-instance interoperability is achieved through DID-based resolution and verifiable claims, allowing seamless operation across federated trust boundaries. For neutrality, the Index will also be hosted at 15 universities to enable resilience and distributed governance.

\subsection{Maintainability}
The logic of the Lean Index Layer is deliberately minimal and stable, reducing operational complexity and surface area for error. Since dynamic state and capabilities are managed through external AgentFacts, updates to the index are infrequent and lightweight.

All records conform to versioned, JSON-LD-based schemas with forward-compatible contexts, ensuring future extensibility. Metadata evolution does not require changes to the Lean Index Layer, enabling rapid iteration at the agent level. Supporting tooling for validation, routing policy, and credential inspection is modular and externally referenceable, ensuring maintainability across long-term, multi-party ecosystems

\begin{table*}[h]
\caption{Comparison of Five Agent Registry Approaches}
\label{tab:agent_regis_comparison}
\centering
\footnotesize
\begin{tabular}{@{}p{2.1cm} p{2.6cm} p{2.6cm} p{2.6cm} p{2.6cm} p{2.6cm}@{}}
\toprule
\textbf{Dimension} & \textbf{MCP} & \textbf{A2A} & \textbf{AGNTCY} & \textbf{Entra Agent ID} & \textbf{NANDA Index} \\
\midrule
Purpose & Centralized publish + discover for MCP servers & Self-hosted capability + endpoint descriptor & Distributed + multi-registry interoperability & Managed enterprise agent directory & Verifiable, privacy-preserving capability facts \\
Discovery Path & REST list + GET by id & Well-known JSON (1 hop) & Semantic discovery based on taxonomies with local and global metasearch & Portal + Graph/Policy APIs & Lean index $\rightarrow$ Facts / PrivateFacts (2 hop) \\
Trust Primitive & GitHub OAuth + DNS TXT & HTTPS + optional token & Sigstore for data provenance attestations, trustless EVM optional. & Azure AD token + policy engine & VC v2 signatures + VC-Status \\
Privacy Option & None (public reads) & None & Notion of private directories and public vs public records. & Directory scope policies & PrivateFactsURL (obfuscated path) \\
Endpoint Freshness & Poll + updated timestamps & Assumed stable (no TTL) & Adaptive DHT cache; digest rev & Platform-managed sync & TTL + rotating / adaptive endpoints \\
Schema Weight & 1--3 KB JSON & 0.3--1 KB JSON & ~4MB + signature & Directory object metadata & 1–3 KB JSON-LD + VC \\
Best Fit & Tool/plugin ecosystems & SaaS-style API agents & Federated / hybrid fleets, multi-registry & Regulated enterprise governance & High-churn, privacy / verifiability critical \\
\bottomrule
\end{tabular}
\end{table*}

\begin{table*}[h]
\caption{Agent Advertisement Models: Card vs. Facts}
\label{tab:agent_card_vs_facts}
\centering
\footnotesize
\begin{tabular}{@{}p{3.2cm} p{4.3cm} p{6.1cm}@{}}
\toprule
\textbf{Feature} & \textbf{A2A Agent Card} & \textbf{NANDA AgentFacts} \\
\midrule
Identifier & Host URL & Stable ID + signed AgentAddr ref \\[4pt]
Endpoints & Single fixed URL & Static + rotating + adaptive (TTL) set \\[4pt]
Capability Model & skills + capabilities.* & skills + capabilities (VC claims) \\[4pt]
Authentication & securitySchemes list & capabilities.authentication + VC issuer attestations \\[4pt]
Integrity Wrapper & Plain JSON over HTTPS & JSON-LD + VC signature + status list \\[4pt]
Privacy Option & None & PrivateFactsURL (dual-path) \\[4pt]
Freshness & HTTP cache only & TTL per endpoint + revocation events \\[4pt]
Revocation & Re-fetch interval & VC-Status (<1s) \\[4pt]
Size (typical) & 0.3–1 KB & 1–3 KB \\[4pt]
Fetch Hops & 1 (well-known) & 2 (Index → Facts) \\[4pt]
Implementation Effort & Low & Medium (VC tooling) \\
\bottomrule
\end{tabular}
\end{table*}

\section{Comparative Evaluation}\label{sec:comparison}

Our comparative analysis is two-fold:

\begin{itemize}
    \item Table~\ref{tab:agent_regis_comparison} compares registry solutions from MCP, A2A (Google), AGNTCY, Entra Agent ID and NANDA Index, analyzing their purpose, discovery paths, trust primitives, privacy mechanisms, endpoint freshness strategies, schema complexity, and best-fit use cases.

    \item Table~\ref{tab:agent_card_vs_facts} provides a detailed feature-level comparison between Agent Cards (used in A2A) and Agent Facts (used in NANDA Index), highlighting key differences—including but not limited to  metadata structure, endpoint modeling, cryptographic guarantees, and extensibility.
\end{itemize}

\section{Evolution of Agent Registry Architectures}\label{sec:phases}

Agent registry systems have evolved from simple, file-based descriptions to distributed, cryptographically-verifiable registries with structured discovery protocols. This section outlines the progression across thtree key phases, each adding layers of interoperability, scalability, trust, and governance. %Understanding these phases helps clarify both the design motivations behind existing systems and the trade-offs made at each maturity level.

\subsection{Static, Isolated Discovery}

The earliest registry mechanisms rely on static files (e.g., JSON or YAML manifests) published at well-known locations on an agent’s domain. These files are primarily consumed manually or by tightly coupled runtimes and contain minimal, static metadata. Common attributes include agent name, endpoint, and basic capabilities.

\textbf{Example:} Google A2A \texttt{/.well-known/agent.json}

\subsection{Dynamic RESTfull APIs }

This phase introduced runtime introspection via HTTP APIs and formally validated JSON schemas. MCP has RESTfull API to find and list available MCP server in the client apps .

\subsection{Verifiable Metadata and Federated Trust and AI agent quilt}

Registries in this phase adopt cryptographic verification and federated trust mechanisms  closer to Nanda Index. Agent metadata is signed using W3C Verifiable Credentials (VCs), PKI certificates, or JSON Canonicalization with hashing and signing. ID-based identities or domain-bound signatures ensure authenticity, while TTL and revocation mechanisms enable fine-grained cache control. These designs enable trust portability, auditability, and agent-to-agent verification. They are well-suited for mobile, privacy-sensitive, or safety-critical deployments. This approach is closer to Nanda Index.

\section{Discussion}\label{sec:conclusion}

The proliferation of autonomous AI agents across enterprise, research, and consumer domains has created an urgent infrastructure challenge: how to discover, identify, and trust agents at Internet scale. This survey examined five representative registry architectures: MCP Metaregistry, A2A, Microsoft Entra Agent ID, NANDA Index, and AGNTCY ADS, each addressing core requirements of security, authentication, scalability, and maintenance through distinct architectural approaches.

Our analysis reveals several key insights that will shape the future of agent discovery infrastructure:

\begin{enumerate}[leftmargin=*]
    \item \textbf{Architectural Trade-offs Are Protocol-Specific.} The "right" registry architecture depends heavily on deployment context. Enterprise environments with existing Azure AD infrastructure benefit from Entra Agent ID's seamless integration and zero-maintenance approach. Open research communities and decentralized applications require the cryptographic guarantees and federated governance of NANDA Index. Protocol-specific ecosystems like MCP benefit from purpose-built registries that leverage existing authentication systems while maintaining simplicity.

    \item \textbf{Decentralization Enables Long-term Sustainability.} While centralized approaches offer operational simplicity, they create single points of failure and vendor lock-in risks that become increasingly problematic as agent ecosystems mature. NANDA Index's federated design and AGNTCY ADS's content-addressed, OCI-aligned model with decentralized rendezvous demonstrate how decentralized architectures can achieve both scalability and community governance. The dual-path resolution pattern in NANDA and ADS's separation of semantic mapping from immutable content address both privacy and integrity concerns.

    \item \textbf{Security Must Be Built-in, Not Bolted-on.} All examined registries recognize that cryptographic integrity is foundational whether through W3C Verifiable Credentials (NANDA), DNS-TXT verification (MCP), or Azure AD's enterprise security controls (Entra). However, AGNTCY ADS stands out with its Sigstore-backed keyless signing and optional DID-based identity, ensuring both data authenticity and server authentication. This approach prevents registry poisoning and enables private discovery patterns.

    \item \textbf{Interoperability Remains the Critical Gap.} Despite architectural differences, these registries serve overlapping use cases and will inevitably need to interoperate as agent ecosystems mature. Cross-protocol discovery, unified namespace management, and portable agent identities represent the next frontier for infrastructure development.

    \item \textbf{Community Governance is Essential for Ecosystem Health.} History shows that the most resilient Internet infrastructure such as DNS, HTTP, and email, emerges from open, multi-stakeholder governance. While proprietary platforms like Entra Agent ID serve specific enterprise needs, the broader agent ecosystem requires community-governed registries that can evolve independently of any single vendor's interests.
\end{enumerate}

\textbf{Implications for Trust-Critical Domains.} 
Healthcare exemplifies why trust primitives matter: a patient care 
coordination agent must verify diagnostic agents' credentials 
(authentication), ensure treatment recommendations haven't been altered 
(integrity), and protect patient privacy during cross-hospital queries 
(confidentiality). NANDA Index's W3C VC signatures and PrivateFactsURL 
support HIPAA-compliant audit trails~\cite{ashrafuzzaman2025blockchain} and privacy-preserving lookups. 
AGNTCY ADS's Sigstore provenance enables "only execute if verified" 
policies required by FDA software-as-medical-device regulations. Entra 
Agent ID's Conditional Access integrates with existing healthcare IAM 
for zero-trust clinical workflows~\cite{li2023trustworthy}. As AI agents expand into regulated 
sectors, registries become the root of trust for safe, accountable 
autonomous systems.

\section{The Path Ahead: A Switchboard for the Agentic Web}
Registry architectures must evolve toward federated models that separate stable identity resolution from dynamic capability metadata, enabling cryptographic trust and cross-organizational discovery without sacrificing privacy or operational autonomy: whether through protocol-specific solutions like MCP's centralized metaregistry and A2A's well-known endpoints, enterprise governance layers like Entra Agent ID, or decentralized approaches combining semantic indexing with content-addressed naming. 

To move forward, our teams are deploying a working solution in two phases, inviting researchers and practitioners to join: Phase 1 establishes the NANDA Index at MIT as a neutral, consortium-governed switchboard for cross-organizational agent coordination; Phase 2 adds semantic discovery and federation capabilities through AGNTCY's Agent Directory, enabling interoperability with heterogeneous registry types via DHT-based routing and OCI-aligned content addressing. 

Just as ICANN enabled the internet to scale through neutral governance and open standards, realizing an Internet of AI Agents requires foundational infrastructure built collaboratively across industry, academia, and civil society, where technical standards, governance frameworks, and reference implementations reflect diverse needs rather than single-vendor control. The decisions made now will shape agent interaction patterns for decades; we invite the community to participate in building this infrastructure openly, ensuring the agentic web serves the entire ecosystem.

\vspace{-5pt}
\bibliographystyle{IEEEtran}
\bibliography{ref}

\end{document}